\documentclass[aps,prb,reprint,floatfix,groupedaddress,superscriptaddress,longbibliography]{revtex4-1}
\usepackage{lipsum}
\usepackage{graphicx}
\usepackage[utf8]{inputenc}
\usepackage{amsmath}
\usepackage{empheq}
\usepackage{geometry}
\usepackage{amssymb}
\usepackage{textcomp}
\usepackage{gensymb}

\DeclareGraphicsExtensions{.pdf,.png}

\begin{document}

\title{Electromagnon dispersion probed by inelastic x-ray scattering}

\author{Sándor Tóth}
\email{sandor.toth@psi.ch}
\affiliation{Laboratory for Neutron Scattering and Imaging, Paul Scherrer Institute, 5232 Villigen PSI, Switzerland}
\affiliation{Laboratory for Quantum Magnetism, EPFL, 1015 Lausanne, Switzerland}

\author{Björn Wehinger}
\affiliation{Laboratory for Neutron Scattering and Imaging, Paul Scherrer Institute, 5232 Villigen PSI, Switzerland}
\affiliation{Department of Quantum Matter Physics, University of Geneva, 1211 Genève, Switzerland}

\author{Katharina Rolfs}
\affiliation{Laboratory for Scientific Developments and Novel Materials, Paul Scherrer Institute, 5232 Villigen PSI, Switzerland}

\author{Turan Birol}
\affiliation{Department of Chemical Engineering and Materials Science, University of Minnesota, 55455 Minneapolis, USA}
\affiliation{Department of Physics and Astronomy, Rutgers University, 08854 New Jersey, USA}

\author{Uwe Stuhr}
\affiliation{Laboratory for Neutron Scattering and Imaging, Paul Scherrer Institute, 5232 Villigen PSI, Switzerland}

\author{Hiroshi Takatsu}
\affiliation{Department of Energy and Hydrocarbon Chemistry, Graduate School of Engineering, Kyoto University, 615-8510 Kyoto, Japan}
\affiliation{Department of Physics, Tokyo Metropolitan University, 192-0397 Tokyo, Japan}

\author{Kenta Kimura}
\affiliation{Division of Materials Physics, Graduate School of Engineering Science, Osaka University, 560-8531 Osaka, Japan}

\author{Tsuyoshi Kimura}
\affiliation{Division of Materials Physics, Graduate School of Engineering Science, Osaka University, 560-8531 Osaka, Japan}

\author{Henrik M. Rønnow}
\affiliation{Laboratory for Quantum Magnetism, EPFL, 1015 Lausanne, Switzerland}

\author{Christian Rüegg}
\affiliation{Laboratory for Neutron Scattering and Imaging, Paul Scherrer Institute, 5232 Villigen PSI, Switzerland}
\affiliation{Department of Quantum Matter Physics, University of Geneva, 1211 Genève, Switzerland}

\date{\textrm{\today}}

\newcommand{\vect}[1]{{\bf{#1}}}
\newcommand{\matr}[1]{\mathsf{#1}}
\newcommand{\licro}{LiCrO$_2$}

\maketitle


\textbf{
Inelastic x-ray scattering with meV energy resolution (IXS) is an ideal tool to measure collective excitations in solids and liquids. In non-resonant scattering condition, the cross section is strongly dominated by lattice vibrations (phonons). However, it is possible to probe additional degrees of freedom such as magnetic fluctuations that are strongly coupled to the phonons. The IXS spectrum of the coupled system will contain not only the phonon dispersion (majority component) but also the so far undetected magnetic correlation function (minority component). Here we report the discovery of strong magnon--phonon coupling in \licro\ that enables the measurement of magnetic correlations throughout the Brillouin-zone via IXS. We found electromagnon excitations and electric dipole active two-magnon excitations in the magnetically ordered phase and paraelectromagnons in the paramagnetic phase of \licro. We predict that the numerous group of (frustrated) magnets with dominant direct exchange and non-collinear magnetism shows similarly strong coupling and surprisingly large and measurable IXS cross section for magnons and multi-magnon processes.
}

The coupling between magnetic and lattice degrees of freedom gives rise to many interesting effects. It can change the ground state properties of the system inducing multiferroic order with ferroelectric polarisation coupled to the magnetic structure \cite{Eerenstein2006,Tokura2006,Cheong2007} or it can generate dynamic mixed magnon--phonon excitations. If the magnon is coupled to a polar phonon, the hybrid excitation is called electromagnon.\cite{Pimenov2006,Sushkov2007,Sushkov2008a} The interest in electromagnons lies in the fact that they enable control of magnetic properties of a material at ultrafast time scales via femtosecond light pulses.\cite{Sheu2014} Moreover optical properties of magnetoelectric materials can be controlled via external DC magnetic field.\cite{Takahashi2011,Kezsmarki2014}
The electric field component of light at the resonant frequency can excite and measure electromagnons only with zero momentum although electromagnons disperse as a function momentum. And as we shall show, maxima in the magnon--phonon coupling can occur at finite momenta inaccessible to zero momentum THz spectroscopy. Inelastic neutron scattering can also identify the magnetic and phononic component of an electromagnon excitation, however previous studies found only small energy shifts of the magnons (majority component) due to magnon--phonon coupling,\cite{Petit2007} while the minority component could not be resolved so far. Here we show that \licro\ is an exceptional material where the magnon--phonon coupling is strong enough to make the minority excitations accessible for inelastic x-ray scattering and thus enables the direct measurement of the electromagnon dispersion.


\begin{figure*}[!htb]
    \centering
	\includegraphics[width=\textwidth]{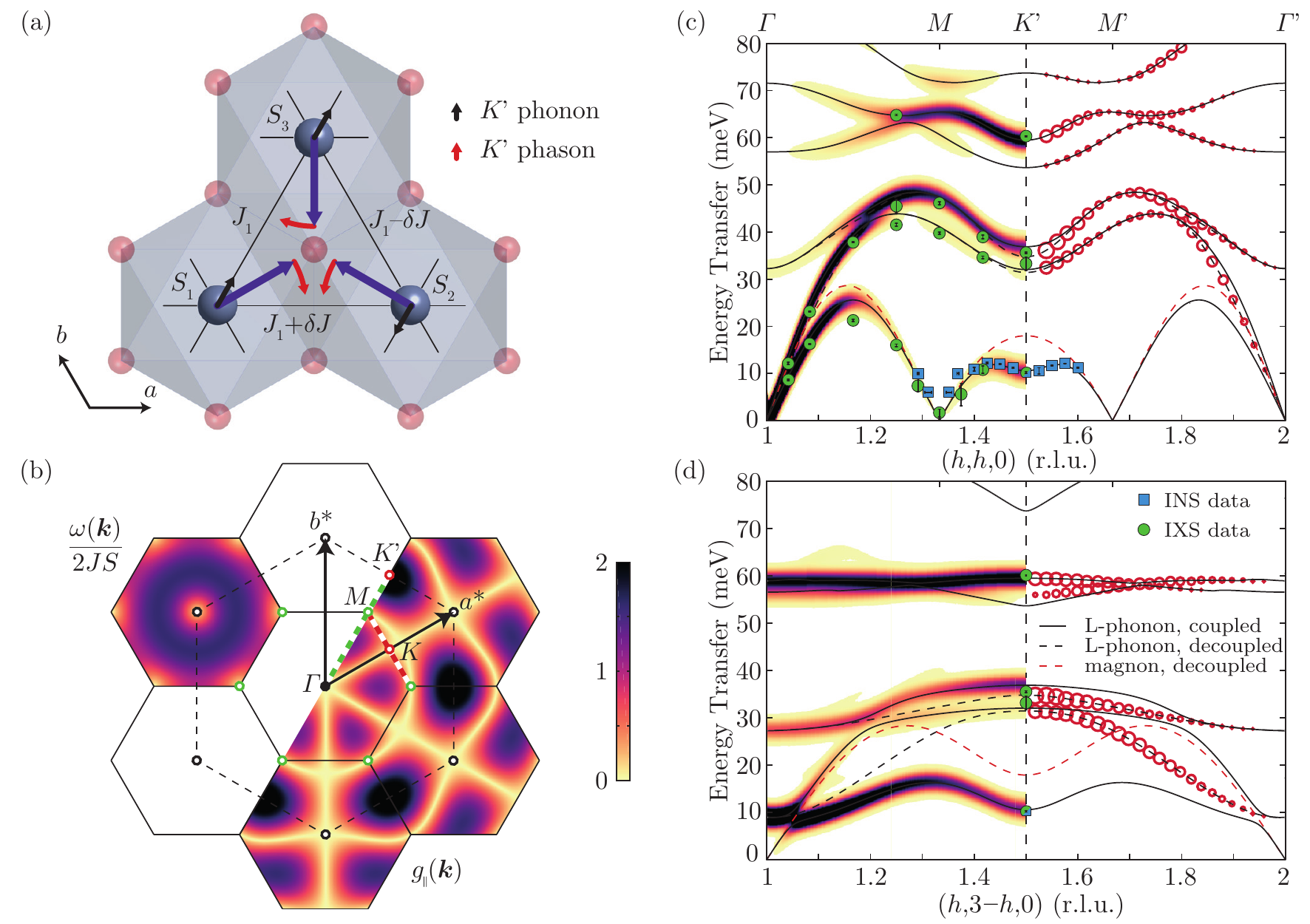}
	\caption{\textbf{Coupled magnon and phonon modes in \licro\ in real and reciprocal space.} (a) A single triangle of Cr$^{3+}$ spins is shown with the surrounding O$_6$ octahedra. Purple arrows depict the helical magnetic structure (rotated into the $ab$-plane for better visibility), the phonon and phason amplitude at the $K'$-point is shown by black and red arrows respectively. (b) Reciprocal space of the triangular lattice with black and dashed hexagons denoting the magnetic and structural Brillouin zones. The upper-left and lower-right colour maps show the phason energy and the $g_\|(\vect{k})$ value respectively (see text). Green and red dashed lines show the path of the IXS and INS measurement, respectively. (c) Comparison of the measured phonon dispersion at 7 K and the coupled magnon--phonon model along the $(h,h,0)$ direction. The colormap on the left half shows the calculated IXS cross section in arbitrary units, while the filled green circles and blue squares with standard deviation denote the measured quasiparticle energies using IXS and INS, respectively. The black dashed and red dashed lines denote the magnon and longitudinal phonon dispersion of the uncoupled model, while the continuous black lines correspond to the coupled dispersion. The empty red circles on the right half denote the $\vect{e}_\lambda\cdot\vect{g}(\vect{k})$ factor that determines the strength of the magnon--phonon coupling for each $\lambda$ phonon mode. (d) Model calculation for in-plane direction perpendicular to $(h,h,0)$.}
\end{figure*}

\licro\ is an excellent realisation of the two-dimensional $S=3/2$ Heisenberg triangular lattice antiferromagnet (TLA) model with only minimal corrections due to structure and symmetry, see Fig.\ 1(a). Dzyaloshinskii-Moriya interactions are forbidden on all bonds due to the space group symmetry of $R\overline{3}m$. Single ion anisotropy is expected to be small due to the octahedral coordination of the Cr$^{3+}$ ion which have half-filled $t_{2g}$ shells resulting in quenched orbital angular momentum. The interplane interactions are weak due to the large separation of the triangular layers. 
\licro\ develops long range magnetic order at $T_N = 61.2$ K.\cite{Sugiyama2009} The magnetic structure is an $ac$-plane helical order with wave vectors of $\vect{k}_m=(1/3,1/3,0)$ and $\vect{k}_m=(-2/3,1/3,1/2)$. The angles between neighbouring spins on the triangular planes are exactly 120\degree\ and the chirality is staggered along the $c$-axis as a result of the double-$Q$ structure.\cite{Kadowaki1995} The magnetic order of a single triangular unit in the $ab$-plane is shown in Fig.\ 1(a). Although high resolution x-ray diffraction does not reveal any symmetry lowering at the magnetic ordering temperature (see Supplementary Fig.\ S1), the staggered chirality implies the appearance of a small symmetry breaking term in the spin Hamiltonian below $T_N$.
The magnetic interactions in the plane are dominated by direct exchange.\cite{Mazin2007a} These interactions are sensitive to the modulation of the bond length, similarly to other Cr$^{3+}$ compounds with short bonds such as ZnCr$_2$O$_4$ \cite{Sushkov2005,Ji2009} and MgCr$_2$O$_4$ \cite{Tchernyshyov2002}. \licro\ furthermore shows a pronounced anomaly in the dielectric constant at $T_N$ but no ferroelectric polarisation could be observed \cite{Seki2008} pointing towards an antiferroelectric ground state induced by the staggered chirality of the triangular layers.\cite{Arima2007,Soda2009} 

\begin{figure}[!htb]
    \centering
	\includegraphics[width=\columnwidth]{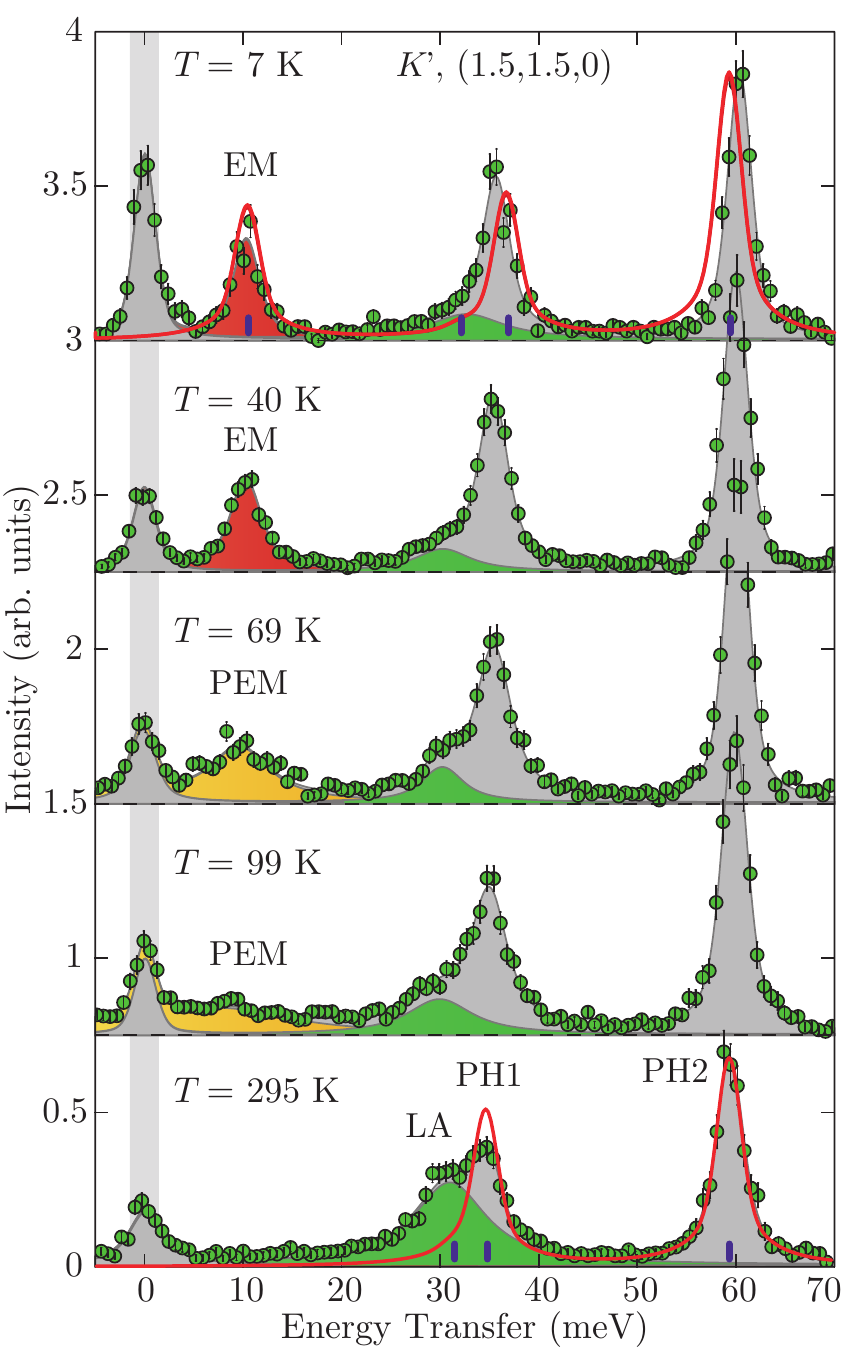}
	\caption{\textbf{Temperature evolution of the measured IXS spectra at the $K'$ point.} Green dots denote experimental data with standard deviation normalised to the same monitor. Red, yellow, green and grey filled areas are fitted electromagnon (EM), paraelectromagnon (PEM), longitudinal acoustic (LA) and optical phonon peaks (PH1, PH2), respectively. Light grey area shows the FWHM of the elastic line. Red lines are the theoretical spectrum of the coupled model at 7 K and the pure phonon model at 295 K. Vertical purple lines show the calculated quasiparticle energies.}
\end{figure}

The IXS spectrum of \licro\ measured at room temperature at $Q = (1.5,1.5,0)$ shows three phonon modes at energies of $30.8(4)$, $34.6(2)$ and $59.2(1)$ meV, see Fig.\ 2. The measured $Q$-point is equivalent to the $K'$ point of the Brillouin zone shifted by $(1,1,0)$, see Fig.\ 1(b). The lowest energy mode has an unusually large intrinsic width of $6.8(5)$ meV (see Methods for details of the data analysis). Upon cooling, the phonon spectrum goes through a dramatic change. The lowest phonon peak looses almost all of its intensity and a new resonance develops gradually below $10$ meV. This new mode appears as a broad diffuse scattering signal at $99$ K, centred at about $8$ meV.
With decreasing temperature the peak becomes more pronounced accompanied with increasing spectral weight and decreasing width. At the lowest measured temperature of $7$ K the peak position is at $10.3(2)$ meV and it has a resolution limited width. The sum of the spectral weight of the lowest phonon mode and the low temperature resonance is independent of temperature, showing that the lowest energy phonon transfers most of its spectral weight to the new mode upon cooling (see Supplementary Fig.\ S5). The measured room temperature phonon spectrum agrees well with the phonon energies determined from \textit{ab initio} calculations (see Methods) shown as vertical purple lines in Fig.\ 2 and the calculated dispersion reveals that the lowest energy observed phonon is a longitudinal acoustic (LA) branch while the two peaks at higher energy correspond to optical branches (PH1 and PH2). While the relative spectral weight of the two upper modes is well reproduced by the calculation, the intensity of the LA phonon is strongly underestimated (see red curve in Fig.\ 2). The calculated energy of the LA phonon is 31.4 meV and its symmetry belongs to the $E_u$ polar representation.

To unambiguously identify the new low energy mode of \licro\ we measured the excitation spectrum by inelastic neutron scattering (INS) along the $(h,1-2h,0)$  reciprocal space direction equivalent to $(h,h,0)$ in the magnetic Brillouin zone, see Supplementary Fig.\ S2. At the $K$ point (equivalent to $K'$) a single spin wave excitation was found in the helical phase at 1.5 K centred at $10.3(1)$ meV. Since neutrons are sensitive to magnetic fluctuations in the measured momentum range, we can conclude that the low energy resonance at $T<T_N$ has not only polar phononic but also magnetic character, thus it is an electromagnon with a finite momentum. Moreover at intermediate temperatures above $T_N$ we call the strongly damped low energy excitation  paraelectromagnon (PEM) due to the lack of both magnetic and electric dipole order. The observed PEM is the phonon coupled excitation of the 2D correlated magnetic ground state, which persists far above $T_N$ due to the low dimensionality of the system. \cite{Kadowaki1995,Moreno2004,Alexander2007}
 


\begin{figure*}[!htb]
    \centering
	\includegraphics[width=\textwidth]{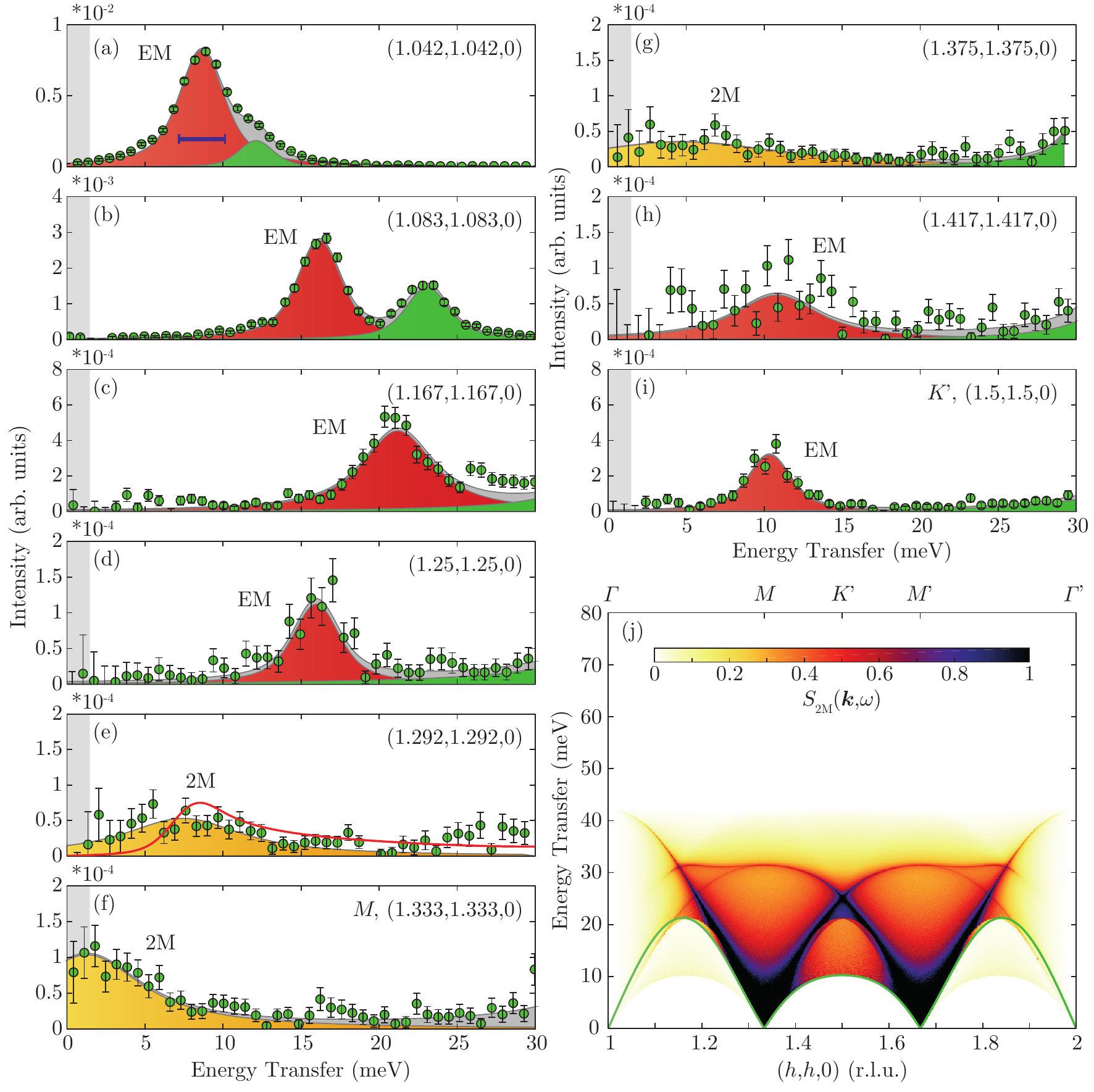}
	\caption{\textbf{IXS electromagnon spectrum measured at $7$ K and two-magnon correlation function.} (a-i) Green dots denote experimental data with standard deviation along $(h,h,0)$ normalised to the same monitor after the subtraction of the elastic peak. Red, yellow, green and grey filled areas are fitted electromagnon (EM), two-magnon (2M), longitudinal acoustic (LA) and optical phonon peaks (PH1, PH2), respectively. Light grey area shows the FWHM of the elastic line. The plots are scaled individually to enhance the visibility of the weak peaks. The red line in (e) shows the theoretical two-magnon spectrum convoluted with the experimental resolution function at $Q=(1.292,1.292,0)$. The horizontal purple bar in (a) shows the FWHM of the instrumental resolution function. (j) The absolute value of the two-magnon correlation function for the pure magnon model. Green line denote the dispersion of the phason spin wave mode.}
\end{figure*}

To determine the coupling mechanism that drives the observed strong magnon--phonon mixing, we measured IXS spectra at multiple points along the $(h,h,0)$ reciprocal space direction at $T=7$ K and fitted the phonon energies. The electromagnon spectrum is reported in Fig.\ 3(a-i). Remarkably, the energy width of the electromagnon excitation increases substantially around the magnetic Bragg point ($M$ point). Since the one magnon excitations are sharp at low temperature, the broad IXS peaks can be due to phonons coupled to the two-magnon (2M) continuum as it is broad for dispersive magnons. For comparison we calculated the two-magnon spectrum for the TLA \cite{Coldea2003} with first and second neighbour antiferromagnetic interactions $J_1 = 8.17$ meV and $J_2=0.556$ meV, shown on Fig.\ 3(j) and a cut at $(1.292,1.292,0)$ in reciprocal space is shown on Fig.\ 3(e). The two-magnon correlations are strongest close to the $M$ point and the centre of the two-magnon spectral weight is expected to be close to the one magnon energy. This explains why the measured electromagnon spectrum continuously changes from a sharp one magnon - one phonon mode to a phonon mixed with the two-magnon continuum as its momentum gets closer to the $M$ point. In the following we model only the single magnon--phonon spectrum. 

The fitted peak positions of both the INS and IXS data are presented in Fig.\ 1(c) together with the model calculations, which will be explained in the following. In general, helical magnetic structures have three spin wave modes: a phason mode with rotation of all spins in the ordering plane and two canting modes correspond to spins canting away from the ordering plane. Strikingly our measured electromagnon spectrum contains only one of the three spin wave modes that according to its dispersion corresponds to the phason mode of the helical magnetic structure. Moreover the two canting modes of the spin spiral are completely decoupled from the phonons. Besides, the phason mode shows a roton like minimum at $K'$. Similar minima were previously observed in several TLAs such as CuCrO$_2$ \cite{Poienar2010,Frontzek2011}, $\alpha$-Cr$_2$O$_4$ \cite{Toth2012} and LuMnO$_3$ \cite{Xiang2011} pointing toward a general sensitivity of the magnon energy at the $K$ point to perturbations such as further neighbour interaction or magnon--phonon coupling.\cite{Kim2007} The electromagnon in \licro\ has large IXS scattering cross section at both the $\Gamma$ and $K'$ points.

The microscopic mechanism that couples the magnons and phonons in \licro\ is the symmetric exchange striction (ES), since the antisymmetric exchange is too weak being a relativistic correction\cite{Moriya1960}. In the following we will show that the measured electromagnon dispersion and IXS cross section can be well described on a single triangular layer assuming strong exchange striction between first neighbour chromium atoms. We will show that in non-collinear magnets ES gives a linear coupling between magnons and phonons thus can generate a strong mixing (for a detailed description see Supplementary Materials). To quantitatively model the spectrum we propose the following Hamiltonian that couples spins to phonons, taking into account the ideal isotropic nature of the spins in \licro:
\begin{align}
	\mathcal{H} = J(r)\sum_{m,n} {\vect{S}}_m\cdot{\vect{S}}_n+\mathcal{H}_\mathrm{L},
	\label{eq:1}
\end{align}
where $J(r)$ is the Heisenberg exchange between first neighbour spins as a function of the bond length $r$, ${\vect{S}}_m$ is the spin vector operator on the $m$th magnetic atom and $\mathcal{H}_\mathrm{L}$ is the Hamiltonian of the lattice vibrations. To simplify Eq.\ \ref{eq:1}, we keep only the constant and linear term from the Taylor expansion of $J(r)$ around the $r_0$ equilibrium bond length. The constant term $J_1$ describes the spin wave dynamics in the absence of phonons, while the linear coefficient $J_\mathrm{mp}$ gives the leading magnon phonon coupling term:
\begin{align}
	\mathcal{H}_\mathrm{mp} = J_\mathrm{mp}\sum_{m,n} \hat{\vect{d}}_{mn}\cdot(\vect{u}_m-\vect{u}_n)\vect{S}_m\cdot\vect{S}_n,
	\label{eq:2}
\end{align}
where $\vect{u}_m$ is the displacement vector of atom $m$ and $\hat{\vect{d}}_{mn}$ is the unit bond vector pointing from atom $m$ to atom $n$. In the magnetically ordered phase if the order is non-collinear $\mathcal{H}_\mathrm{mp}$ linearly couples the phonon and magnon bosonic operators $a_\lambda(\vect{k})$ and $b(\vect{k})$. After applying the linear Holstein-Primakoff approximation and using a rotating coordinate system for the spins \cite{Coldea2003,Chernyshev2009,Toth2014a} the equation simplifies to:
\begin{align}
	\mathcal{H}_\mathrm{mp}=i\sum_{\vect{k},\lambda}\vect{\gamma}_\lambda(\vect{k})a_\lambda(\vect{k})\left(b^\dagger(\vect{k})-b(-\vect{k})\right)+\text{h.c.},
\end{align}
where $\lambda$ indexes the phonon modes. The coupling term $\gamma_\lambda(\vect{k})$ is given by:
\begin{align}
	\gamma_\lambda(\vect{k}) = -\frac{3}{4}J_\mathrm{mp}S\sqrt{\frac{S\hbar}{M\omega_\lambda(\vect{k})}} \vect{e}_\lambda(\vect{k})\cdot\vect{g}(\vect{k}),
\end{align}
where $M$ is the mass of the magnetic atom, $\omega_\lambda(\vect{k})$ and $\vect{e}_\lambda(\vect{k})$ are the energy and amplitude of the $\lambda$ phonon on the chromium atom. The $\vect{g}(\vect{k})$ geometrical factor for the Bravais lattice of magnetic atoms is the following:
\begin{align}
	\vect{g}(\vect{k}) = \sum_\vect{d}\hat{\vect{d}} \sin(2\pi \vect{k}_m\cdot\vect{d})\left[ \cos(2\pi\vect{k}\cdot\vect{d})-1 \right],
\end{align}
where the sum runs through bonds denoted by $\vect{d}$ (where exchange striction is active). The linear coupling vanishes for non-collinear magnetic order, because $\vect{g}(\vect{k})$ is zero for $\vect{k}_m=0$. The coupled model can be solved using Bogoliubov transformation and the corresponding neutron and x-ray scattering cross sections can be calculated (see Supplementary Materials). The inactivity of the additional two canting spin wave modes of the helical structure in the IXS spectrum can be also explained within our model. These modes mathematically are related to a coordinate transformation turning the spin spiral into a ferromagnet having a single spin wave mode (the phason mode). Transforming back into the lab coordinate system explains the two additional modes (same dispersion with $\pm\vect{k}_m$ momentum shift) that can be observed by inelastic neutron scattering. However the exchange striction is isotropic in spin space therefore the transformation does not produce additional observable electromagnon modes.

For a full interpretation of the IXS spectrum of \licro, we start with the pure phonon spectrum in the paramagnetic phase determined from \textit{ab initio} calculations. The dispersion of the longitudinal phonons are shown in Fig.\ 1(c-d) by black dashed lines and the full phonon spectrum in Supplementary Fig.\ S8. 
The calculated dispersion relation already agrees well with the measured phonon energies showing that the magnon--phonon coupling introduces only minor energy shifts. The introduction of $J_\mathrm{mp}$ will mix the phason and phonon amplitudes. The strongest mixing is calculated to be between the longitudinal (LA) and transverse (TA) acoustic phonon branches of the 2D triangular planes and the phason spin wave mode of the helical magnetic structure. The wave vector dependent intensity of the IXS electromagnon signal is proportional to $g_\|(\vect{k}) = \vect{g}(\vect{k})\cdot\hat{\vect{k}}$ which is largest along the $(h,h,0)$ in reciprocal space and zero at lattice and magnetic Bragg points, see Fig.\ 1(b). It is important to note that although $g_\|(\vect{k})$ is zero at the $K$-point, the coupled dispersion is the same as at $K'$ just both $\vect{g}(\vect{k})$ and $\vect{e}(\vect{k})$ vectors are rotated by $90^\circ$ thus invisible for IXS. The largest mixing amplitude is expected at $\Gamma$ and $K'$ in agreement with our experimental results. Impressively, the strong coupling causes a roton minimum of the spin wave dispersion at $K'$ downwards renormalising the phason energy by 42\% even though the lowest phonon mode is $20$ meV higher in energy.

To determine the parameters of the coupled model, we fitted the experimental electromagnon dispersions using $J_1$ and $J_\mathrm{mp}$ as parameters. The best model parameters are $J_1=6.00(25)$ meV and $J_\mathrm{mp}=65(4)$ meV/\AA. If fitted, an additional second neighbour exchange interaction in the triangular planes is zero within error bar. The optimised model Hamiltonian describes both the coupled dispersion (see black lines in Fig.\ 1(c)) and the IXS cross section (see red line in Fig.\ 2) very well. Some deviation close to the $\Gamma$ point is due to the overestimation of the speed of sound from the \textit{ab initio} method. The real space dynamics of the strong coupling at the $K'$ point is visualised in Fig.\ 1(a). At this wave vector the longitudinal acoustic phonon (black arrows) shortens and lengthens the $S_1-S_2$ and $S_2-S_3$ bonds, respectively. The excited phason mode is in phase with the phonon that makes the $S_1-S_2$ bond stronger ($J_1+\delta J$) while the $S_2-S_3$ bond weaker ($J_1+\delta J$). Thus a ferromagnetic fluctuation on the longer bond and antiferromagnetic fluctuation on the shorter bond is energetically favourable if exchange striction is present explaining the reduction of the phason energy and the roton minimum at the $K'$ point. The $S_1-S_2$ bond is inactive at this wave vector since it changes neither length nor relative spin orientation. The strongest electromagnon cross section is expected close to the $\Gamma$ point even though the $\mathcal{H}_\mathrm{mp}$ coupling term vanishes at $\Gamma$. This is due to the decreasing energy separation between the LA phonon and the phason mode towards the zone centre. 

In conclusion, we reported inelastic x-ray scattering data on \licro\ that revealed an electromagnon that is the phason mode of the helical spin order coupled to a longitudinal acoustic phonon. We identified the exchange striction between first neighbour chromium ions as the microscopic coupling mechanism. Fitting the model parameters to the measured electromagnon dispersion we could reproduce both the experimental dispersion and the dynamical structure factor for inelastic x-ray scattering. Beside the one magnon process we also found signature of coupling between the acoustic phonon branches and the two-magnon continuum around the magnetic Bragg points that can be explained by including higher order corrections to our linear theory. In the paramagnetic phase we observed for the first time paraelectromagnon excitations, a heavily damped electromagnon stabilised by the low dimensional magnetic correlations of the 2D triangular lattice. By accessing the momentum dependence, our results shed light on a much richer physics of electromagnons that is beyond the reach of THz light experiments. The reported measurement also shows how inelastic x-ray scattering can be used to probe magnetic correlations with high energy and momentum resolution in certain systems.
This study will open a route towards measuring magnetic correlations at extreme conditions using diamond anvil cells. Indeed, IXS can be performed with samples as thin as 10-20 $\mu$m, which allow extending such studies up to Mbar pressure.\cite{Antonangeli2004} It is furthermore possible to work with evanescent wave fields in grazing angle conditions which allows surface sensitive studies, measurements on thin films and multilayer systems.\cite{Murphy2005,Serrano2011}

\section{Methods}

\subsection{Crystal growth}
\licro\ single crystals for INS were grown by the Li$_2$O-B$_2$O$_3$ flux or Li$_2$O-PbO-B$_2$O$_3$ flux methods for IXS and INS measurements respectively. A typical growth was done by a mixture of Li$_2$O, Cr$_2$O$_3$, and B$_2$O$_3$ or with additional PbO. The mixture was heated at 1300\degree C and then slowly cooled down to 800\degree C or 900\degree C respectively.

\subsection{Inelastic x-ray scattering}
Inelastic x-ray scattering was measured on the ID28 beamline at the European Synchrotron Radiation Facility (ESRF) along the reciprocal space direction $(h,h,0)$ at temperatures $295$, $99$, $69$, $40$ and $7$ K using incident photon energy of $17.794$ keV ($\lambda = 0.6968$ \AA) and beam size of $50\times 50$ $\mu$m$^2$. Since the sample was a thin plate perpendicular to $(0,0,1)$, we choose the $(h,h,l)$ scattering plane to minimise absorption. The ID28 instrumental energy resolution has a pseudo-Voigt profile with $2.71(2)$ meV and $3.3(1)$ meV full width at half maximum (FWHM) of the Gaussian and Lorentzian components and a mixing parameter of 0.63(2). The constant momentum transfer scans were fitted with a line shape that is the instrumental energy resolution convoluted with a Lorentzian (all intrinsic width is given by the FWHM of the Lorentzian in the main text) that models the finite lifetime of the excitations. The momentum resolution of the ID28 spectrometer is close to rectangular with $0.027$ \AA$^{-1}$ and $0.076$ \AA$^{-1}$ horizontal and vertical width perpendicular to the momentum transfer, while the longitudinal momentum resolution is at least two orders of magnitude better than the transverse. 

\subsection{Inelastic neutron scattering}
Inelastic neutron scattering was measured on the EIGER triple-axis spectrometer at SINQ at the Paul Scherrer Institut (PSI) using fixed final neutron energy of 14.7 meV, double focusing graphite monochromator and horizontal focusing graphite analyser. To eliminate spurious scattering a pyrolytic graphite filter was applied after the sample. We have used a $50$ mg single crystal of \licro\ and performed measurements at $1.5$ K. Due to the small sample size, the spin wave signal was only collected close to the magnetic Bragg points along the $(h,1-2h,0)$ direction and at $(1/2,1/2,0)$ in reciprocal space. The spin wave peak as a function of neutron energy transfer was fitted with a Gaussian function. 

\subsection{Phonon calculation}
Lattice dynamics calculations were performed using the finite displacement method within density functional theory \cite{Parlinski1997}. Distorted atomic configurations were generated and the induced forces of a $x \times x \times x $ supercell were computed by total energy calculations using Projector Augmented Waves method as implemented in VASP \cite{PAW,VASP3,VASP2}. A shifted $8\times 8 \times 8$ k-point grid is used for the ionic relaxations and the calculation of Born Effective charges by perturbation theory\cite{DFPT} in the primitive unit cell. While the internal ionic coordinates are relaxed, the lattice constant is kept fixed to the experimental value in order to reduce the error due to unit cell volume. The valence electrons treated explicitly by the VASP PAW potentials are $1s^22s^2sp^1$ for Li, $3p^63d^54s^1$ for Cr, and $2s^22p^4$ for O. A plane wave cutoff of 500 eV, which is 25\% larger than suggested, is used and tested to provide good convergence. \texttt{PBEsol} exchange correlation functional\cite{PBE,PBEsol} is employed for all calculations. In order to account for the underestimation of on-site correlations by GGA, the DFT+U approximation\cite{DFTU} is used with a U of 3 eV which has previously been shown to faithfully reproduce the spin-phonon properties of Cr oxides in the same implementation.\cite{Wysocki2016} Dynamical matrices throughout the Brillouin zone were computed using Fourier transformation as implemented in Phonopy \cite{phonopy} and non-analytical term corrections due to finite Born charges were applied. A shifted $4\times4\times4$ $k$-point grid has been used for sampling the electronic structure of the primitive unit cell.

\subsection{Magnon--phonon coupled model}
The spin wave model and the coupled magnon--phonon model was solved numerically using a modified version of SpinW \cite{Toth2014a}.

\section{Author contributions}
S.\ T.\ and B.\ W.\ carried out inelastic x-ray spectroscopy and analysed data. S.\ T., K.\ R.\ and U.\ S.\ carried out inelastic neutron scattering experiments and analysed data. K.\ R., H.\ T., K.\ K.\ and T.\ K.\ synthesised samples. B.\ W.\ and T.\ B.\ carried out \textit{ab initio} calculations. The results were discussed and interpreted by S.\ T.\ and Ch.\ R. The manuscript was written by S.\ T.\ with input from all the authors.

\section{Acknowledgement}
We thank Andrea Scaramucci and Michel Kenzelmann for helpful discussions on electromagnons. We acknowledge the European Synchrotron Radiation Facility for provision of synchrotron radiation facilities and we would like to thank Thanh-Tra Nguyen for assistance in using beamline ID28 and Christina Drathen for collecting data on beamline ID22. We also thank Céline Besnard for preliminary x-ray diffraction measurements. This work is based on experiments performed at the Swiss spallation neutron source SINQ, Paul Scherrer Institute, Villigen, Switzerland. The research leading to these results has received funding from the European Community's Seventh Framework Programme (FP7/2007-2013) under Grant Agreement No. 290605  (COFUND: PSI-FELLOW). K.\ K.\ and T.\ K.\ were supported by JSPS KAKENHI (grant no. 24244058).
\section{Competing Interests}
The authors declare that they have no competing financial interests.
\section{Correspondence}
Correspondence and requests for materials should be addressed to S. Tóth~(email:sandor.toth@psi.ch).

\bibliographystyle{naturemag_noURL}

\begin{thebibliography}{10}
\expandafter\ifx\csname url\endcsname\relax
  \def\url#1{\texttt{#1}}\fi
\expandafter\ifx\csname urlprefix\endcsname\relax\def\urlprefix{URL }\fi
\providecommand{\bibinfo}[2]{#2}
\providecommand{\eprint}[2][]{\url{#2}}

\bibitem{Eerenstein2006}
\bibinfo{author}{Eerenstein, W.}, \bibinfo{author}{Mathur, N.~D.} \&
  \bibinfo{author}{Scott, J.~F.}
\newblock \bibinfo{title}{{Multiferroic and magnetoelectric materials}}.
\newblock \emph{\bibinfo{journal}{Nature}} \textbf{\bibinfo{volume}{442}},
  \bibinfo{pages}{759} (\bibinfo{year}{2006}).

\bibitem{Tokura2006}
\bibinfo{author}{Tokura, Y.}
\newblock \bibinfo{title}{{Multiferroics as Quantum Electromagnets}}.
\newblock \emph{\bibinfo{journal}{Science (80-. ).}}
  \textbf{\bibinfo{volume}{312}}, \bibinfo{pages}{1481} (\bibinfo{year}{2006}).

\bibitem{Cheong2007}
\bibinfo{author}{Cheong, S.-W.} \& \bibinfo{author}{Mostovoy, M.}
\newblock \bibinfo{title}{{Multiferroics: a magnetic twist for
  ferroelectricity.}}
\newblock \emph{\bibinfo{journal}{Nat. Mater.}} \textbf{\bibinfo{volume}{6}},
  \bibinfo{pages}{13} (\bibinfo{year}{2007}).

\bibitem{Pimenov2006}
\bibinfo{author}{Pimenov, A.} \emph{et~al.}
\newblock \bibinfo{title}{{Possible evidence for electromagnons in multiferroic
  manganites}}.
\newblock \emph{\bibinfo{journal}{Nat. Phys.}} \textbf{\bibinfo{volume}{2}},
  \bibinfo{pages}{97} (\bibinfo{year}{2006}).

\bibitem{Sushkov2007}
\bibinfo{author}{Sushkov, A.~B.}, \bibinfo{author}{{Vald{\'{e}}s Aguilar}, R.},
  \bibinfo{author}{Park, S.}, \bibinfo{author}{Cheong, S.-W.} \&
  \bibinfo{author}{Drew, H.~D.}
\newblock \bibinfo{title}{{Electromagnons in Multiferroic YMn$_2$O$_5$ and
  TbMn$_2$O$_5$}}.
\newblock \emph{\bibinfo{journal}{Phys. Rev. Lett.}}
  \textbf{\bibinfo{volume}{98}}, \bibinfo{pages}{027202}
  (\bibinfo{year}{2007}).

\bibitem{Sushkov2008a}
\bibinfo{author}{Sushkov, A.~B.}, \bibinfo{author}{Mostovoy, M.},
  \bibinfo{author}{{Vald{\'{e}}s Aguilar}, R.}, \bibinfo{author}{Cheong, S.-W.}
  \& \bibinfo{author}{Drew, H.~D.}
\newblock \bibinfo{title}{{Electromagnons in multiferroic RMn$_2$O$_5$
  compounds and their microscopic origin}}.
\newblock \emph{\bibinfo{journal}{J. Phys. Condens. Matter}}
  \textbf{\bibinfo{volume}{20}}, \bibinfo{pages}{434210}
  (\bibinfo{year}{2008}).
\newblock arXiv:\eprint{0806.1207}.

\bibitem{Sheu2014}
\bibinfo{author}{Sheu, Y.~M.} \emph{et~al.}
\newblock \bibinfo{title}{{Using ultrashort optical pulses to couple
  ferroelectric and ferromagnetic order in an oxide heterostructure}}.
\newblock \emph{\bibinfo{journal}{Nat. Commun.}} \textbf{\bibinfo{volume}{5}},
  \bibinfo{pages}{5832} (\bibinfo{year}{2014}).

\bibitem{Takahashi2011}
\bibinfo{author}{Takahashi, Y.}, \bibinfo{author}{Shimano, R.},
  \bibinfo{author}{Kaneko, Y.}, \bibinfo{author}{Murakawa, H.} \&
  \bibinfo{author}{Tokura, Y.}
\newblock \bibinfo{title}{{Magnetoelectric resonance with electromagnons in a
  perovskite helimagnet}}.
\newblock \emph{\bibinfo{journal}{Nat. Phys.}} \textbf{\bibinfo{volume}{8}},
  \bibinfo{pages}{121} (\bibinfo{year}{2011}).

\bibitem{Kezsmarki2014}
\bibinfo{author}{K{\'{e}}zsm{\'{a}}rki, I.} \emph{et~al.}
\newblock \bibinfo{title}{{One-way transparency of four-coloured spin-wave
  excitations in multiferroic materials}}.
\newblock \emph{\bibinfo{journal}{Nat. Commun.}} \textbf{\bibinfo{volume}{5}},
  \bibinfo{pages}{3203} (\bibinfo{year}{2014}).

\bibitem{Petit2007}
\bibinfo{author}{Petit, S.} \emph{et~al.}
\newblock \bibinfo{title}{{Spin Phonon Coupling in Hexagonal Multiferroic
  YMnO$_3$}}.
\newblock \emph{\bibinfo{journal}{Phys. Rev. Lett.}}
  \textbf{\bibinfo{volume}{99}}, \bibinfo{pages}{266604}
  (\bibinfo{year}{2007}).

\bibitem{Sugiyama2009}
\bibinfo{author}{Sugiyama, J.} \emph{et~al.}
\newblock \bibinfo{title}{{$\mu$+SR investigation of local magnetic order in
  LiCrO$_2$}}.
\newblock \emph{\bibinfo{journal}{Phys. Rev. B}} \textbf{\bibinfo{volume}{79}},
  \bibinfo{pages}{184411} (\bibinfo{year}{2009}).

\bibitem{Kadowaki1995}
\bibinfo{author}{Kadowaki, H.}, \bibinfo{author}{Takei, H.} \&
  \bibinfo{author}{Motoya, K.}
\newblock \bibinfo{title}{{Double-$Q$ 120\degree\ structure in the Heisenberg
  antiferromagnet on rhombohedrally stacked triangular lattice LiCrO$_2$}}.
\newblock \emph{\bibinfo{journal}{J. Phys. Condens. Matter}}
  \textbf{\bibinfo{volume}{7}}, \bibinfo{pages}{6869} (\bibinfo{year}{1995}).

\bibitem{Mazin2007a}
\bibinfo{author}{Mazin, I.}
\newblock \bibinfo{title}{{Electronic structure and magnetism in the frustrated
  antiferromagnet LiCrO$_2$: First-principles calculations}}.
\newblock \emph{\bibinfo{journal}{Phys. Rev. B}} \textbf{\bibinfo{volume}{75}},
  \bibinfo{pages}{094407} (\bibinfo{year}{2007}).

\bibitem{Sushkov2005}
\bibinfo{author}{Sushkov, A.~B.}, \bibinfo{author}{Tchernyshyov, O.},
  \bibinfo{author}{{Ratcliff II}, W.}, \bibinfo{author}{Cheong, S.-W.} \&
  \bibinfo{author}{Drew, H.~D.}
\newblock \bibinfo{title}{{Probing Spin Correlations with Phonons in the
  Strongly Frustrated Magnet ZnCr$_2$O$_4$}}.
\newblock \emph{\bibinfo{journal}{Phys. Rev. Lett.}}
  \textbf{\bibinfo{volume}{94}}, \bibinfo{pages}{137202}
  (\bibinfo{year}{2005}).

\bibitem{Ji2009}
\bibinfo{author}{Ji, S.} \emph{et~al.}
\newblock \bibinfo{title}{{Spin-Lattice Order in Frustrated ZnCr$_2$O$_4$}}.
\newblock \emph{\bibinfo{journal}{Phys. Rev. Lett.}}
  \textbf{\bibinfo{volume}{103}}, \bibinfo{pages}{037201}
  (\bibinfo{year}{2009}).

\bibitem{Tchernyshyov2002}
\bibinfo{author}{Tchernyshyov, O.}, \bibinfo{author}{Moessner, R.} \&
  \bibinfo{author}{Sondhi, S.}
\newblock \bibinfo{title}{{Order by Distortion and String Modes in Pyrochlore
  Antiferromagnets}}.
\newblock \emph{\bibinfo{journal}{Phys. Rev. Lett.}}
  \textbf{\bibinfo{volume}{88}}, \bibinfo{pages}{067203}
  (\bibinfo{year}{2002}).

\bibitem{Seki2008}
\bibinfo{author}{Seki, S.}, \bibinfo{author}{Onose, Y.} \&
  \bibinfo{author}{Tokura, Y.}
\newblock \bibinfo{title}{{Spin-Driven Ferroelectricity in Triangular Lattice
  Antiferromagnets ACrO$_{2}$ (A=Cu, Ag, Li, or Na)}}.
\newblock \emph{\bibinfo{journal}{Phys. Rev. Lett.}}
  \textbf{\bibinfo{volume}{101}}, \bibinfo{pages}{067204}
  (\bibinfo{year}{2008}).

\bibitem{Arima2007}
\bibinfo{author}{Arima, T.-h.}
\newblock \bibinfo{title}{{Ferroelectricity Induced by Proper-Screw Type
  Magnetic Order}}.
\newblock \emph{\bibinfo{journal}{J. Phys. Soc. Japan}}
  \textbf{\bibinfo{volume}{76}}, \bibinfo{pages}{073702}
  (\bibinfo{year}{2007}).

\bibitem{Soda2009}
\bibinfo{author}{Soda, M.}, \bibinfo{author}{Kimura, K.},
  \bibinfo{author}{Kimura, T.}, \bibinfo{author}{Matsuura, M.} \&
  \bibinfo{author}{Hirota, K.}
\newblock \bibinfo{title}{{Electric Control of Spin Helicity in Multiferroic
  Triangular Lattice Antiferromagnet CuCrO$_2$ with Proper-Screw Order}}.
\newblock \emph{\bibinfo{journal}{J. Phys. Soc. Japan}}
  \textbf{\bibinfo{volume}{78}}, \bibinfo{pages}{124703}
  (\bibinfo{year}{2009}).

\bibitem{Moreno2004}
\bibinfo{author}{Moreno, N.} \emph{et~al.}
\newblock \bibinfo{title}{{Magnetic properties of the frustrated
  antiferromagnet LiCrO$_2$}}.
\newblock \emph{\bibinfo{journal}{J. Magn. Magn. Mater.}}
  \textbf{\bibinfo{volume}{272-276}}, \bibinfo{pages}{E1023}
  (\bibinfo{year}{2004}).

\bibitem{Alexander2007}
\bibinfo{author}{Alexander, L.~K.}, \bibinfo{author}{B{\"{u}}ttgen, N.},
  \bibinfo{author}{Nath, R.}, \bibinfo{author}{Mahajan, A.~V.} \&
  \bibinfo{author}{Loidl, A.}
\newblock \bibinfo{title}{{$^7$Li NMR studies on the triangular lattice system
  LiCrO$_2$}}.
\newblock \emph{\bibinfo{journal}{Phys. Rev. B}} \textbf{\bibinfo{volume}{76}},
  \bibinfo{pages}{064429} (\bibinfo{year}{2007}).

\bibitem{Coldea2003}
\bibinfo{author}{Coldea, R.}, \bibinfo{author}{Tennant, D.} \&
  \bibinfo{author}{Tylczynski, Z.}
\newblock \bibinfo{title}{{Extended scattering continua characteristic of spin
  fractionalization in the two-dimensional frustrated quantum magnet
  Cs$_2$CuCl$_4$ observed by neutron scattering}}.
\newblock \emph{\bibinfo{journal}{Phys. Rev. B}} \textbf{\bibinfo{volume}{68}},
  \bibinfo{pages}{134424} (\bibinfo{year}{2003}).

\bibitem{Poienar2010}
\bibinfo{author}{Poienar, M.}, \bibinfo{author}{Damay, F.},
  \bibinfo{author}{Martin, C.}, \bibinfo{author}{Robert, J.} \&
  \bibinfo{author}{Petit, S.}
\newblock \bibinfo{title}{{Spin dynamics in the geometrically frustrated
  multiferroic CuCrO$_2$}}.
\newblock \emph{\bibinfo{journal}{Phys. Rev. B}} \textbf{\bibinfo{volume}{81}},
  \bibinfo{pages}{104411} (\bibinfo{year}{2010}).

\bibitem{Frontzek2011}
\bibinfo{author}{Frontzek, M.} \emph{et~al.}
\newblock \bibinfo{title}{{Magnetic excitations in the geometric frustrated
  multiferroic CuCrO$_2$}}.
\newblock \emph{\bibinfo{journal}{Phys. Rev. B}} \textbf{\bibinfo{volume}{84}},
  \bibinfo{pages}{094448} (\bibinfo{year}{2011}).

\bibitem{Toth2012}
\bibinfo{author}{Toth, S.} \emph{et~al.}
\newblock \bibinfo{title}{{Magnetic Soft Modes in the Distorted Triangular
  Antiferromagnet $\alpha$-CaCr$_2$O$_4$}}.
\newblock \emph{\bibinfo{journal}{Phys. Rev. Lett.}}
  \textbf{\bibinfo{volume}{109}}, \bibinfo{pages}{127203}
  (\bibinfo{year}{2012}).

\bibitem{Xiang2011}
\bibinfo{author}{Xiang, H.~J.}, \bibinfo{author}{Kan, E.~J.},
  \bibinfo{author}{Zhang, Y.}, \bibinfo{author}{Whangbo, M.~H.} \&
  \bibinfo{author}{Gong, X.~G.}
\newblock \bibinfo{title}{{General theory for the ferroelectric polarization
  induced by spin-spiral order}}.
\newblock \emph{\bibinfo{journal}{Phys. Rev. Lett.}}
  \textbf{\bibinfo{volume}{107}}, \bibinfo{pages}{157202}
  (\bibinfo{year}{2011}).

\bibitem{Kim2007}
\bibinfo{author}{Kim, J.} \& \bibinfo{author}{Han, J.}
\newblock \bibinfo{title}{{Coupling of phonons and spin waves in a triangular
  antiferromagnet}}.
\newblock \emph{\bibinfo{journal}{Phys. Rev. B}} \textbf{\bibinfo{volume}{76}},
  \bibinfo{pages}{054431} (\bibinfo{year}{2007}).

\bibitem{Moriya1960}
\bibinfo{author}{Moriya, T.}
\newblock \bibinfo{title}{{Anisotropic Superexchange Interaction and Weak
  Ferromagnetism}}.
\newblock \emph{\bibinfo{journal}{Phys. Rev.}} \textbf{\bibinfo{volume}{120}},
  \bibinfo{pages}{91--98} (\bibinfo{year}{1960}).

\bibitem{Chernyshev2009}
\bibinfo{author}{Chernyshev, A.} \& \bibinfo{author}{Zhitomirsky, M.~E.}
\newblock \bibinfo{title}{{Spin waves in a triangular lattice antiferromagnet:
  Decays, spectrum renormalization, and singularities}}.
\newblock \emph{\bibinfo{journal}{Phys. Rev. B}} \textbf{\bibinfo{volume}{79}},
  \bibinfo{pages}{144416} (\bibinfo{year}{2009}).

\bibitem{Toth2014a}
\bibinfo{author}{Toth, S.} \& \bibinfo{author}{Lake, B.}
\newblock \bibinfo{title}{{Linear spin wave theory for single-Q incommensurate
  magnetic structures}}.
\newblock \emph{\bibinfo{journal}{J. Phys. Condens. Matter}}
  \textbf{\bibinfo{volume}{27}}, \bibinfo{pages}{166002}
  (\bibinfo{year}{2015}).

\bibitem{Antonangeli2004}
\bibinfo{author}{Antonangeli, D.} \emph{et~al.}
\newblock \bibinfo{title}{{Elasticity of Cobalt at High Pressure Studied by
  Inelastic X-Ray Scattering}}.
\newblock \emph{\bibinfo{journal}{Phys. Rev. Lett.}}
  \textbf{\bibinfo{volume}{93}}, \bibinfo{pages}{215505}
  (\bibinfo{year}{2004}).

\bibitem{Murphy2005}
\bibinfo{author}{Murphy, B.~M.} \emph{et~al.}
\newblock \bibinfo{title}{{Phonon Modes at the
  $2H\mathrm{\text{-}}{\mathrm{NbSe}}_{2}$ Surface Observed by Grazing
  Incidence Inelastic X-Ray Scattering}}.
\newblock \emph{\bibinfo{journal}{Phys. Rev. Lett.}}
  \textbf{\bibinfo{volume}{95}}, \bibinfo{pages}{256104}
  (\bibinfo{year}{2005}).

\bibitem{Serrano2011}
\bibinfo{author}{Serrano, J.} \emph{et~al.}
\newblock \bibinfo{title}{{InN Thin Film Lattice Dynamics by Grazing Incidence
  Inelastic X-Ray Scattering}}.
\newblock \emph{\bibinfo{journal}{Phys. Rev. Lett.}}
  \textbf{\bibinfo{volume}{106}}, \bibinfo{pages}{205501}
  (\bibinfo{year}{2011}).

\bibitem{Parlinski1997}
\bibinfo{author}{Parlinski, K.}, \bibinfo{author}{Li, Z.~Q.} \&
  \bibinfo{author}{Kawazoe, Y.}
\newblock \bibinfo{title}{{First-Principles Determination of the Soft Mode in
  Cubic ZrO$_2$}}.
\newblock \emph{\bibinfo{journal}{Phys. Rev. Lett.}}
  \textbf{\bibinfo{volume}{78}}, \bibinfo{pages}{4063} (\bibinfo{year}{1997}).

\bibitem{PAW}
\bibinfo{author}{Bl{\"{o}}chl, P.~E.}
\newblock \bibinfo{title}{{Projector augmented-wave method}}.
\newblock \emph{\bibinfo{journal}{Phys. Rev. B}} \textbf{\bibinfo{volume}{50}},
  \bibinfo{pages}{17953} (\bibinfo{year}{1994}).

\bibitem{VASP3}
\bibinfo{author}{Kresse, G.} \& \bibinfo{author}{Furthm{\"{u}}ller, J.}
\newblock \bibinfo{title}{{Efficiency of ab-initio total energy calculations
  for metals and semiconductors using a plane-wave basis set}}.
\newblock \emph{\bibinfo{journal}{Comput. Mater. Sci.}}
  \textbf{\bibinfo{volume}{6}}, \bibinfo{pages}{15} (\bibinfo{year}{1996}).

\bibitem{VASP2}
\bibinfo{author}{Kresse, G.} \& \bibinfo{author}{Furthm{\"{u}}ller, J.}
\newblock \bibinfo{title}{{Efficient iterative schemes for ab initio
  total-energy calculations using a plane-wave basis set}}.
\newblock \emph{\bibinfo{journal}{Phys. Rev. B}} \textbf{\bibinfo{volume}{54}},
  \bibinfo{pages}{11169} (\bibinfo{year}{1996}).

\bibitem{DFPT}
\bibinfo{author}{Baroni, S.}, \bibinfo{author}{de~Gironcoli, S.},
  \bibinfo{author}{{Dal Corso}, A.} \& \bibinfo{author}{Giannozzi, P.}
\newblock \bibinfo{title}{{Phonons and related crystal properties from
  density-functional perturbation theory}}.
\newblock \emph{\bibinfo{journal}{Rev. Mod. Phys.}}
  \textbf{\bibinfo{volume}{73}}, \bibinfo{pages}{515--562}
  (\bibinfo{year}{2001}).

\bibitem{PBE}
\bibinfo{author}{Perdew, J.~P.}, \bibinfo{author}{Burke, K.} \&
  \bibinfo{author}{Ernzerhof, M.}
\newblock \bibinfo{title}{{Generalized Gradient Approximation Made Simple}}.
\newblock \emph{\bibinfo{journal}{Phys. Rev. Lett.}}
  \textbf{\bibinfo{volume}{77}}, \bibinfo{pages}{3865--3868}
  (\bibinfo{year}{1996}).

\bibitem{PBEsol}
\bibinfo{author}{Perdew, J.~P.} \emph{et~al.}
\newblock \bibinfo{title}{{Restoring the Density-Gradient Expansion for
  Exchange in Solids and Surfaces}}.
\newblock \emph{\bibinfo{journal}{Phys. Rev. Lett.}}
  \textbf{\bibinfo{volume}{100}}, \bibinfo{pages}{136406}
  (\bibinfo{year}{2008}).

\bibitem{DFTU}
\bibinfo{author}{Liechtenstein, A.~I.}, \bibinfo{author}{Anisimov, V.~I.} \&
  \bibinfo{author}{Zaanen, J.}
\newblock \bibinfo{title}{{Density-functional theory and strong interactions:
  Orbital ordering in Mott-Hubbard insulators}}.
\newblock \emph{\bibinfo{journal}{Phys. Rev. B}} \textbf{\bibinfo{volume}{52}},
  \bibinfo{pages}{R5467} (\bibinfo{year}{1995}).

\bibitem{Wysocki2016}
\bibinfo{author}{Wysocki, A.~L.} \& \bibinfo{author}{Birol, T.}
\newblock \bibinfo{title}{{Magnetically-induced phonon splitting in
  ACr$_2$O$_4$ spinels from first principles}}.
\newblock \emph{\bibinfo{journal}{Phys. Rev. B}} \textbf{\bibinfo{volume}{93}},
  \bibinfo{pages}{134425} (\bibinfo{year}{2015}).

\bibitem{phonopy}
\bibinfo{author}{Togo, A.} \& \bibinfo{author}{Tanaka, I.}
\newblock \bibinfo{title}{{First principles phonon calculations in materials
  science}}.
\newblock \emph{\bibinfo{journal}{Scr. Mater.}} \textbf{\bibinfo{volume}{108}},
  \bibinfo{pages}{1} (\bibinfo{year}{2015}).

\end{thebibliography}

\end{document}